# Background Rates in Several Hard X-Ray Photon Counting Pixel Array Detectors

Alfred Q. R. Baron[†] and Daisuke Ishikawa

*Materials Dynamics Laboratory, RIKEN SPring-8 Center, 1-1-1 Kouto, Sayo, Hyogo 679-5148, JAPAN*
*Precision Spectroscopy Division, CSRR, SPring-8/JASRI, 1-1-1 Kouto, Sayo, Hyogo 679-5198, JAPAN*

**Background rates in several pixel array detectors are investigated with an eye toward using them with hard, >6 keV, x-rays in very low-rate experiments - e.g., at signal rates <0.1/s/cm$^2$. Measured background event rates for a detector with an unshielded 0.75 mm thick CdTe sensor on the experimental floor at SPring-8 varied from 0.4 /s/cm$^2$ with a 6 keV lower-level discriminator (LLD) threshold to 0.2/s/cm$^2$ with a 75 keV LLD threshold.  The background for a detector with a 1 mm thick silicon sensor was smaller, ~0.08/s/cm$^2$ for a 3keV threshold dropping to ~0.07/s/cm$^2$ at a 17 keV threshold.  These rates are dominated by terrestrial sources, such as gamma rays emitted from trace impurities in concrete, with only a small contribution, <0.01/s/cm$^2$, from direct detection of cosmic ray muons (CRMs).  15 mm of Pb shielding reduces the measured rates to < 0.05/s/cm$^2$ in CdTe and to < 0.02/s/cm$^2$ in 1 mm silicon.  Additional processing, "time slicing", as may be used in low-rate experiments where backgrounds may be most problematic, is shown to reduce the background rates further, to < 0.004/s/cm$^2$ with the silicon sensor, and to between 0.002 and 0.02/s/cm$^2$ for CdTe, where the exact value for CdTe depends sensitively on the detector threshold, and the use of the retriggering and/or the use of dual discriminator thresholds.  Background rates are presented as functions of discriminator threshold, shielding, and processing.  We also discuss the magnitude of, and the correction for, the limitation of the detector dynamic range that can be introduced by the time slicing.  Finally, we present one example where time slicing was used.**

## 1. Introduction:

Position sensitive pixel array detectors (PADs) offer many advantages for x-ray scattering experiments, allowing data collection to generally be much more efficient, and, in fact, changing the way one thinks about doing experiments (see e.g. [1] and reference therein).  This is especially true for diffraction experiments where typical rates are relatively high, usually >> 1/s/cm$^2$ for the regions of interest.  However, it can also be desirable to use position sensitive detectors in very low-rate measurements, where, for the purposes of this discussion we are considering rates that might be ~0.01/s/cm$^2$ of detecting area, or for a ~0.1 mm scale pixel, ~10$^{-6}$/s/pixel.  In particular, we have an interest pushing the limits of inelastic x-ray scattering (IXS) experiments, so we carefully explore the backgrounds in the present paper.  As a point of comparison, the detectors we now use for IXS are CdZnTe chips custom made by Hamamatsu, with a single element size of 2x2x1 mm$^3$. The background rate for these detectors, used in vacuum with a discriminator window (both a lower and an upper level) is typically ~10$^{-3}$/s at 21.7 keV, or ~0.025/s/cm$^2$.  We would like to achieve a similar background rate with a PAD and, when considering ambitious experiments, would be happy to do better, with desired background rates at the level of 0.001/s/cm$^2$.

A naïve estimate of the background rates suggests achieving the desired level, ~0.001/s/cm$^2$, *might* be straightforward. The electronic (dark) noise in pixel array systems used for hard x-rays is negligible, so the main limit for low-rate experiments is expected to be real background events.  Assuming sufficient shielding has been arranged, and no internal sources, one might assume the remaining background originates from the flux of cosmic ray muons (CRMs), as these tend to penetrate most shielding.  Being charged particles, muons will tend to leave a track in the detector.  While this means that each muon passing through the sensor can illuminate (deposit charge in) a large number of pixels as it passes through them, it also suggests that these tracks might be readily identified and processed out.  If one assumes such processing is done, then one may estimate the unavoidable background rate from the direct cosmic ray muon flux (about 1/minute/cm$^2$ at sea level) and the expected probability of a muon depositing charge in just one pixel (<5% given the rather high pixel aspect ratio for the detectors used here) as <0.05/minute/cm$^2$ or  < 0.001/s/cm$^2$.  However, the observed rates are much larger.  This highlights the importance of measuring the background rate.

---
[†] baron@spring8.or.jp

The organization of this paper is as follows. First, we describe the detectors tested and mention some properties of the Si and CdTe materials used in the detector sensors as are relevant for understanding the detector response to x-rays. Then we discuss some relevant properties of cosmic ray muons at the earth's surface. The measured background rates are then presented for 3 detectors both with and without lead shielding. Then there is a brief discussion of different background events and the pixel number distributions of background events. We discuss time slicing to reduce backgrounds in low-rate experiments, showing its efficacy for several cases, and the magnitude of, and correction for, systematic errors introduced in the processing. Finally, we present an example of a high-resolution inelastic x-ray scattering experiment where the background rate is a notable issue and where the benefits of the processing can be clearly observed.

Before continuing further, we show several images to help illustrate the background problem we are investigating. Figure 1 shows the background measured for 2000s for several specific cases, including just directly placing the detector on a

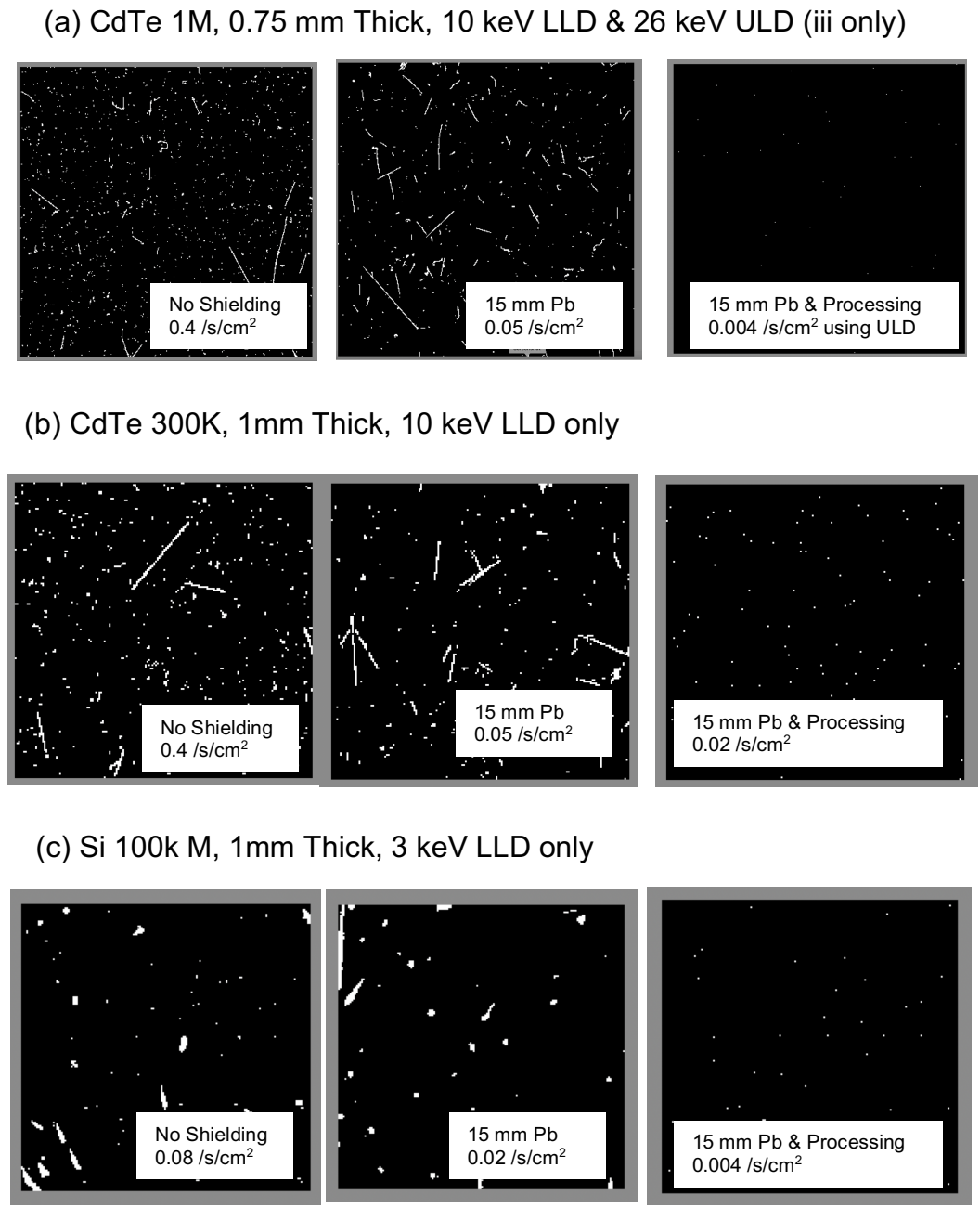

**Figure 1.** A 30x30 mm$^2$ area from each of the detectors investigated, integrated over 2000 s with LLDs set as indicated. The detectors were measured just as placed on the experimental floor (left column), then with 15 mm Pb shielding (center). The Pb shielding clearly reduces the number of few-pixel events. The third (right-most) panel in each row shows the same data as the 15mm Pb case, but with "time-slicing" used to remove events that are identified as background, as discussed in section 9 of the main text. The rates indicated in each plot are for the number of events/s/cm$^2$ where an "event" is a non-empty frame collected in 0.001s. We emphasize, this number is *not* pixels illuminated, and *not* counts in the detector. Given that frame rates were high enough so that the number of exposed frames was typically <1% of the total, these event rates are reasonable indicators of the real background rate – i.e.: multiple events in one frame only occur about in <1.2% of the frames with events and do not impact the event rate for the precision given. The events are more difficult to see in (a) iii as compared to (c) iii because the pixel size is smaller for the 1M detector. Also, the blotchy character of the large events in (c) is reduced – the events become more track-like - if the LLD is increased.

table on the experimental floor with a lower-level discriminator (LLD) threshold as indicated, then as measured with 15 mm of Pb shielding added to reduce the background gamma background radiation, and then with processing, as discussed below, to remove background events that may be clearly differentiated from signal events. In the first two cases, the tracks from CRM events are clearly visible, and the main effect of the Pb shielding is to remove many of the smaller (few pixel) events. Finally, instead of summing all the frames as was done for the second column, the third column shows the result when the data from the measurement with the Pb shielding is time sliced, allowing tracks to be removed and significantly reducing the background rate.

## 2. Detectors Investigated

We present data from 3 detectors made by DECTRIS [2] as listed in table 1. The Pilatus3 detectors have 1 mm thick sensors with a 0.172 mm pixel pitch. One of these had a silicon sensor, so was most efficient below 20 keV while the other had a CdTe sensor. The Eiger2 detector had a CdTe, sensor, 0.75 mm thick, and a smaller, 0.075 mm pixel pitch. All detectors were covered by aluminized mylar windows with a flow of dry nitrogen around the detector, and a water circulator was used to stabilize the detector near room temperature.

| Type | Si 100K - M | CdTe 300K | CdTe 1M |
|---|---|---|---|
| Series | PILATUS3 X 100K-M | PILATUS3 X CdTe 300K | EIGER2 X CdTe 1M |
| Sensor | 1 mm Silicon | 1 mm CdTe | 0.75 mm CdTe |
| Outer Area | 8.38 x 3.35 = 28.1 cm$^2$ | 8.38x10.65= 89.3 cm$^2$ | 7.71 x 7.97 = 61.4 cm$^2$ |
| Active Area | 28.1 cm$^2$ | 82.6 cm$^2$ | 59.1 cm$^2$ |
| Pixel Pitch | 0.172 mm | 0.172 mm | 0.075 mm |
| Nominal Active Pixels | 487x195 = 94965 | 3x16x60x97 279360 | 4x513x512 1,050,624 |
| Masked Pixels | 1 | 96 | 36 |
| Delivery Year | 2020 | 2019 | 2022 |

**Table I. Detectors Investigated**

In the simplest case, a photon counting PAD detector needs a lower-level discriminator (LLD) to distinguish signal events from electronic noise. Often the LLD is set to half of the detected x-ray energy to make a reasonable first approximation for properly counting events that share charge between pixels: if the charge is shared between two pixels, then setting the threshold at half the x-ray energy will generally allow one count to appear in one pixel but prevents a second pixel registering a count. Of course, if the charge is shared between more than two pixels, or if there are fluorescence events, as is important for CdTe, the situation becomes more complicated. The Pilatus3 detectors have one threshold, while the Eiger2 has two, allowing parallel collection of two images with different thresholds. The silicon detector, being designed primarily for lower energy x-rays has thresholds that can be varied from 2.7 to 18 keV while the Pilatus3 CdTe 300K has a threshold variable from 8 to 40 keV and the Eiger2 between 4 and <80 keV. For the Pilatus3 300K, the number of hot pixels increased when the threshold was set either below 10 keV or at 35 keV so most measurements with that detector were done from LLD thresholds of 10 to 30 keV (nominal operating energies of 20 to 60 keV). For the Eiger2 CdTe 1M no such issues existed, but we mainly focused on LLD thresholds <40 keV consistent with planned uses for that detector.

In addition to a lower level threshold, all the detectors also have an "instant-retriggering system" – similar to a time-over-threshold multiplier - that is designed to preserve linearity to higher rates [3,4]. This also acts as a crude integrator or upper-level discriminator: sufficiently large events lead to more than one count in a pixel, and so can be readily recognized (in a low-rate limit) as non-x-ray events. These multi-count single hit events were bigger (had more counts) in the CdTe detectors than in the silicon but did appear in both. They appeared more frequently in the Eiger2 detector, as might be related to a change in circuit design. All measurements, unless otherwise specified, were made with the retriggering active as multicount events in a single pixel could then be identified as background and removed. This was most effective for the Eiger2 detector.

## 3. Detector Sensor Materials

We investigated detectors with sensors made out of Si and CdTe. The basic properties of the materials are summarized in table 1. One notes that the presence of the K edges and fluorescence lines in the hard x-ray region for the CdTe make the response notably different than Si. This can be seen in the calculated interaction or attenuation lengths in figure 2 [5] but also has broader impact [6,7] as we will discuss.

Silicon, with its very low K-edge energy (<2 keV) has a relatively simple response. Fluorescence events are both unlikely (<5% fluorescence yield), and absorbed without travelling very far, thus the great majority of absorbed x-rays result in a relatively short (~10-micron scale) fast electron track, leading to a local deposition of energy (at least until Compton scattering becomes significant at higher energy). However, the disadvantage of silicon is that the attenuation length of x-rays in silicon is already on the mm scale at 20 keV, and quickly increases at higher energy – see figure 1. Thus the efficiency[‡] of silicon sensors, which are typically not more than 1 mm thick and often less, drops off quickly with increasing x-ray energy, with, e.g., measured the efficiency for a thick, 1 mm, silicon detector being about 62% at 20 keV but dropping to about 22% at 30 keV and to about 10% at 40 keV [8]. While one might consider a grazing incidence geometry to increase the path length in the silicon this leads both to a reduction in the effective (projected) active area of the detector and a significant loss in position resolution when the attenuation length is large, as one x-ray trajectory is spread over many pixels, so the effective position resolution will approach the thickness of the sensor element.

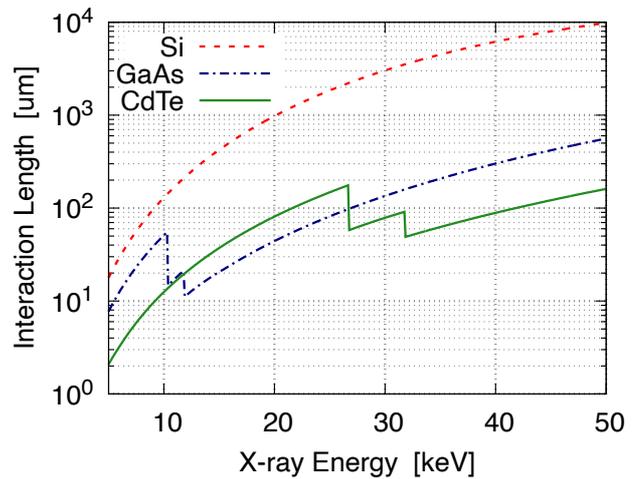

**Figure 2.** Calculated (1/e) attenuation or interaction length, including photoelectric, Compton, and Rayleigh scattering. Based on [5]

CdTe offers much better x-ray stopping power than silicon, but this comes at a cost of a complex response when one exceeds the K-edges at 26.7 keV for Cd and 31.8 keV for Te. K-holes in these materials have a fluorescence yield of 85% or more so most absorbed x-rays at higher energies will tend to lead to fluorescence. This complicates the response on multiple levels (see [7]). On the one hand, there will be an escape peak at the incident energy minus the K-shell fluorescence energy, which, depending on discriminator thresholds, may reduce the efficiency. On another, the attenuation length for the fluorescence x-rays is rather long: ~ 120 microns for the Cd $K_\alpha$ fluorescence, and ~ 60 microns for Te $K_\alpha$ fluorescence. These lengths are comparable to the transverse dimensions of the detector pixels, so it is easy for an incident x-ray generate a K-shell hole in one pixel, which then leads to a fluorescence event that is detected in a different pixel. These concerns account for the rather complex pulse height distribution in these detectors at high energy [7]. They also tend to reduce efficiency when the detector threshold is set at half the incident energy: while one would naively estimate >90% efficiencies up to 50 keV based on the stopping power of a 0.75 mm thick CdTe sensor, the measured values of the efficiency for uniform illumination over several pixels is ~80% for a 0.15 mm pixel, and ~70% for a 0.075mm pixel between 25 and 50 keV when the discriminator threshold is set at half the x-ray energy [7]. However, even so, at energies above ~25 keV, the efficiency of CdTe is a significant improvement compared to even rather thick, 1 mm, silicon.

|  | **Si** | **CdTe** |
|---|---|---|
| **Density** | 2.32 g/cc | 5.85 g/cc |
| **Muon Energy Deposition** (4 GeV Muon) | 1.9 MeV-cm$^2$/g  44 keV/100 μm | 1.6 MeV-cm$^2$/g  94 keV/100 μm |
| **Edges (keV)** | K=1.84 | Cd: K=26.71 ; $L_1$ = 4.02  Te: K = 31.81 ; $L_1$: 4.94 |
| **K Fluorescence Yield** | 4.8% | Cd: 85%   Te: 88% |
| **Fluorescence Lines (keV)** | $K_{\alpha 1}$: 1.74  $K_{\alpha 2}$: 1.74  $K_{\beta 3}$: 1.83 | Cd: $K_{\alpha 1}$ = 23.2  $K_{\alpha 2}$ = 23.0  $K_{\beta 1}$=26.1  $K_{\beta 2}$=26.1  Te: $K_{\alpha 1}$: 27.5  $K_{\alpha 2}$: 27.2  $K_{\beta 1}$: 31.0  $K_{\beta 3}$: 31.0 |

**Table II. Sensor Material Properties**

### 4. Cosmic Ray Muons (CRMs)
It is useful to review properties of the cosmic ray muon (CRM) background [9,10]. At sea-level, the flux (to within about 20%) is 1 muon per minute per cm$^2$ of horizontal detecting area, or 0.017/s/cm$^2$. This is without shielding, so can be reduced if, for example, the detector is placed underground or otherwise in a very heavily shielded environment (see also [11]). The distribution of muon trajectories peaks at zenith, and falls off roughly as $\cos^2(\theta_z)$, where $\theta_z$ is the angular deviation of the muon trajectory from zenith ($\theta_z \leq \pi/2$), though there is also a correlation between the angular distribution and muon energy. Considering the $\cos^2(\theta_z)$ angular dependence, and neglecting the angle-energy correlation, the direct CRM rate in a vertically oriented detector (all detectors used were vertical, as is often the case for SR facilities) will be reduced by a factor of 2 to about 0.008 /s/cm$^2$ of vertical detecting area. The energy of the muons is distributed over a large range [9] – but the mean is near 4 GeV which is just above the minimum ionizing particle energy in most materials [12]. The average energy deposition of a 4 GeV cosmic ray muon is about 1.9 MeV cm$^2$/g for Si and about 1.6 MeV cm$^2$/g for CdTe [12], leading to an expected average deposition of about 44 keV/100 μm in silicon (2.33 g/cc) and about 94 keV/100 μm in CdTe (5.85 g/cc). These are nominal values, and the real energy deposition will be broadly distributed due to variation in muon energies, varying path lengths through the pixels, and fluctuations in energy loss for a given path length (Landau distribution). However, noting the pixel sizes are 172 μm square by 1 mm thick, or 75 μm square by 0.75 mm thick, one can expect that cosmic ray muons passing through a pixel will often - in fact, usually, for lower thresholds - deposit an amount of energy that is above the lower threshold setting of a detector when it is used for hard x-rays.

---
[‡] We use "efficiency" to mean the probability that an incident x-ray generates a pulse over threshold in the detector.

## 5. Other Background Sources

Another source of background radiation (see, e.g., chapter 20 of [10], and [11]) is the decay of unstable nuclei – this includes radon, $^{222}$Rn, in the atmosphere, and also $^{40}$K, $^{232}$Th, $^{238}$U as can be present in trace amounts in many materials, including concrete. The various decay paths lead to a large number of possible gamma ray lines [13] and the impact of these is *not* negligible. For example, in reference [14] a 3" NaI detector without shielding in the lab was reported to have a background rate of ~500/minute/cm$^2$ for photon energies > 100 keV, of which most events deposited < 400 keV in the detector. Given that the interaction probability for a 200 keV photon passing through 1 mm of material is ~3% for Si and ~17% for CdTe, it is then likely that CRMs at a rate of ~1/minute/cm$^2$ will *not* dominate the observed background event rate for the detectors used here, if the detectors remain unshielded.

## 6. Measured Background Rates

Figure 3 shows the measured background event[§] rates in the CdTe Eiger2 1M detector as a function of LLD setting. We collected and processed dual threshold data with that detector at a 1 kHz frame rate. The relatively short, 0.001 s, acquisition time insured that only a small fraction of the frames (<1.2%) had events in them. Dividing the fraction of the frames with events in them by the 0.001 s duration of each frame and the active area of the detector then provides a good estimate of the background event rate as is shown in figure 3. The rate monotonically decreases with threshold, as one would expect. However, the rate without shielding is much higher (25 to 50 times, depending on threshold) the expected direct detection of CRM events, indicating that a significant number of events are either from indirect detection of CRMs (e.g., muons impacting the detector back-plane) or from terrestrial sources. The latter is confirmed by using Pb shielding: placing 5 mm of Pb above, in front and to the sides of the detector, configuration (1), reduced the rate by a factor of 2, while adding another sheet below the detector, configuration (2), reduced the rate by another factor of 2. On the one hand, the reduction from 5 mm of Pb is much larger than the expected reduction of GeV CRMs, and, on the other, if the main source were CRMs, which are already well shielded from below by the earth, then one would expect no significant change in rate between configurations (1) and (2). Thus the results suggest the main source of background is not CRMs, but, probably, gamma radiation from terrestrial sources, such as trace amounts of decaying isotopes in the concrete floor of the experimental hall, *etc*. as is consistent also with the result in [14] mentioned in the preceding section. Finally, the detector was surrounded by 3x5mm thick lead plates on all sides (except the back where the cables exited) and placed on a table with a 50 mm thick steel top – configuration (3). This shows the background rate can be reduced to < ~0.05/events/cm$^2$ in CdTe with shielding. While this rate is still higher than we desire, it is also possible to consider processing to differentiate between signal and background, as is discussed in section 9, below.
Figure 4 shows the measured rates for the Pilatus3 based detectors with the 1 mm thick CdTe and Si sensors. The event rates in the 1 mm thick CdTe are very similar to those for the 0.75 mm thick CdTe Eiger2, above, but slightly larger consistent with the increased thickness. The silicon detector has a lower (approximately 1/5) background rate compared to the CdTe detectors as is qualitatively consistent with the background source being predominantly high-energy (>100 keV) gammas from radioactive decay, for which the probability of interacting in the Si is much less than in CdTe. When a 15 mm Pb shield is used, the background rate in silicon is within about a factor of two of the floor expected from CRMs directly impacting the sensor.

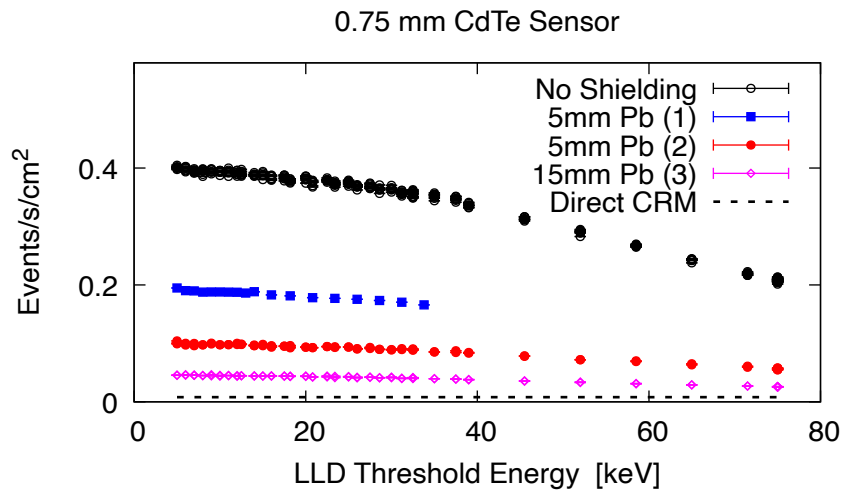

**Figure 3.** Measured background rates as a function of discriminator threshold for the top 29.6 cm$^2$ of an Eiger 2 detector with a 0.75 mm thick CdTe sensor. An "event" is any frame (1ms exposure) having non-zero counts. Each point is the result of processing 2 to 10 million frames at a 1 kHz frame rate. The rates are as indicated for no shielding, for (1) 5 mm Pb shielding in front, to the sides and above the detector and (2) as for (1) but also below the detector, and (3) for 15 mm Pb all around the detector. Also shown is the estimated rate for passage of CRMs through the detector (0.008 /s/cm$^2$). See the text for further discussion.

---

[§] In this work we generally focus on *event* rates where one event is counted whenever a frame (or some area of a frame) has non-zero counts, without regard for the number of counts or (except as explicitly discussed) the number of pixels. This is because while this event rate is reasonably stable and reproducible, the count rate or even the rate of illuminated pixels suffers huge fluctuations that are dominated by the variable trajectories of CRMs through the detector, which sometimes illuminate just a few pixels, but sometimes can illuminate hundreds of pixels (see figures 5 and 6).

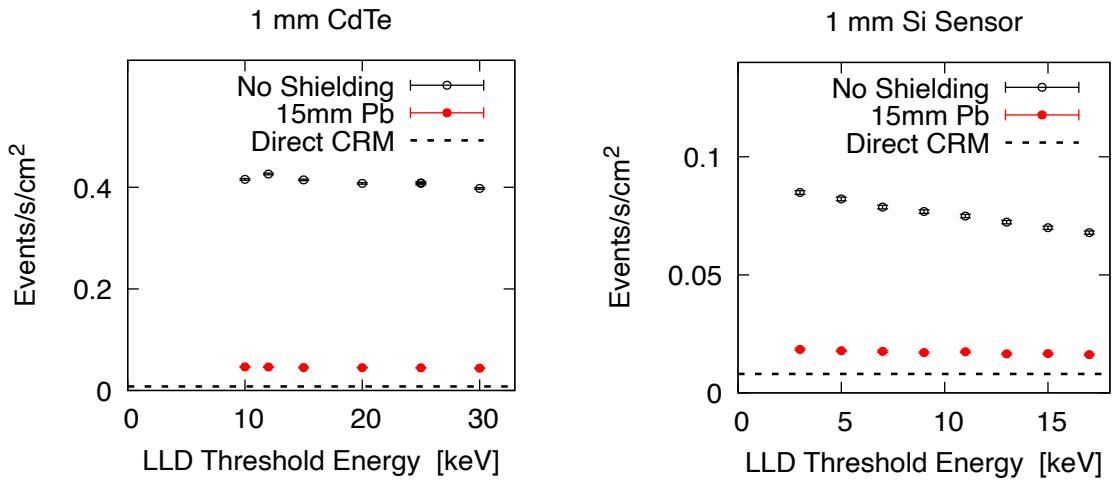

**Figure 4.** Measured background rates as a function of discriminator threshold for the two Pilatus3 detectors. See also the caption for figure 3.

## 7. Illuminated Pixels per Event

As a starting point for considering additional processing to reduce the background we investigate the number of pixels illuminated by each background event. The distributions of illuminated pixels for several runs with 0.001 s frames (0.005s for the Si detector), with and without shielding, is shown in figure 5 on both logarithmic and liner scales (inset). For all detectors the rate falls quickly as the pixel number increases, but, on the log scale, is seen to extend to high values

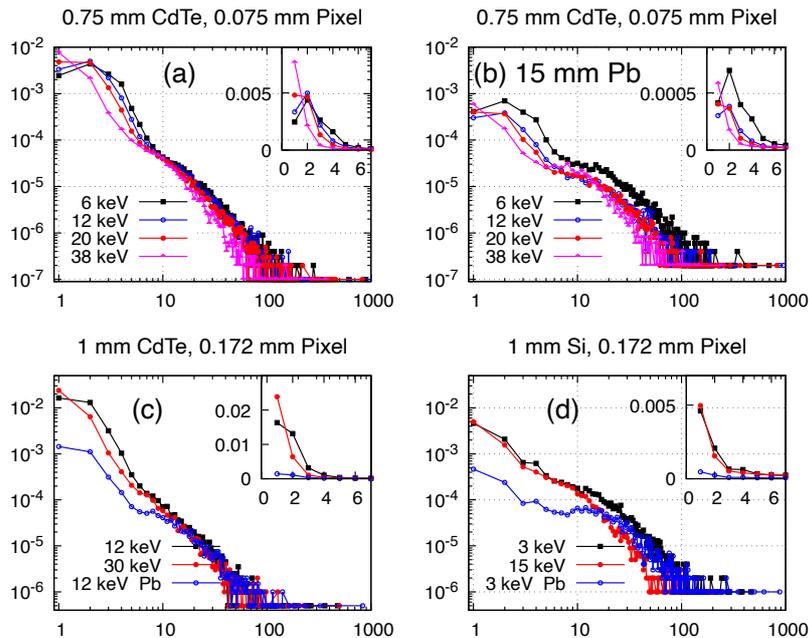

**Figure 5.** Distribution of the number of illuminated pixels in background events for the indicated conditions. (a,b) for the CdTe Eiger2 1M-B, (c) CdTe Pilatus3 300k and (d) Si Pilatus3 100k-M. In all cases the x axis is the number of illuminated pixels and the y axis is the fraction of frames with that number of pixels with counts.

(corresponding to tracks from CRM events – see the next section). However, as is apparent, many, and for lower thresholds, even most, of the background events illuminate more than a single pixel. Thus, there is immediately some hope that by selectively discarding multi-pixel events, one may significantly reduce the effective background rate. One difference between the detector responses is that for the smaller, 0.075 mm, pixel CdTe Eiger2 at lower thresholds, the most likely number of pixels illuminated is 2, whereas for all the other detectors it is 1. That is probably due to a combination of an increased chance of events with charge sharing as the pixel size is smaller and an increased chance the fluorescence events can travel to the next pixel. The primary impact of the Pb shielding is a reduction in the fraction of frames that have <5-10 illuminated pixels. This is consistent with the high pixel number events being CRM tracks, as GeV-scale CRMs are relatively unaffected by 15 mm Pb, while mid-energy (100-500 keV) gammas would be significantly reduced.

**8. Large Background Events**

We investigated large background events in the CdTe Eiger2 detector. One can plot various correlations in the data, and we found that to discover patterns it was useful to consider a 3-D correlation between illuminated pixels, total counts and the ratio between the highest count pixel and the next highest count pixel in a given frame. We remind the reader that the retriggering was active so that the number of counts in a pixel can be >1. While the retriggering system is mainly designed to improve linearity at higher rates, it was also useful here, as it means the counts in a pixel roughly correlates with the total energy deposited: with the retriggering active, occasional background events could have to 10's, 100's or even 1000's of counts registered in a pixel. Investigating the correlations, and looking at many frames, allowed us to identify 3 main types of larger background events, as are shown in figure 6:

1. "Crater" events had the largest number of counts. These are characterized by roughly symmetric (aspect ratio ~1) group of illuminated pixels over an area of < 1 mm$^2$ where typically one or two of the central pixels had very few (1-30) counts but were surrounded by a ring of pixels with very large counts (100-3000) and then finally an outer border with some small (1-2) counts. Thus, these

(a) "Crater" Events

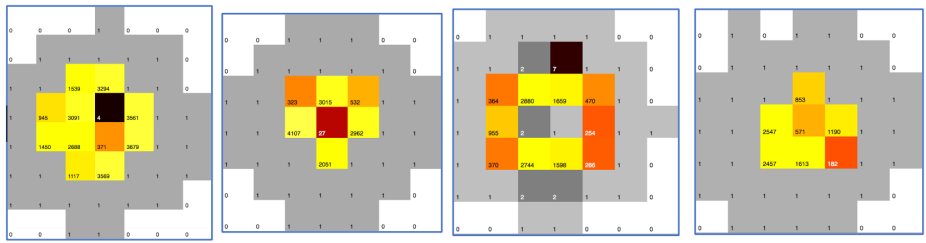

(b) "Cluster" Events

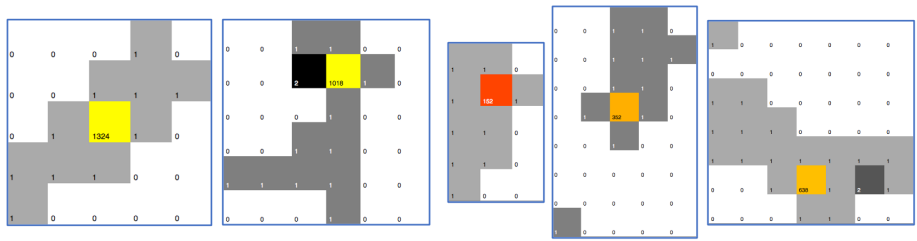

(c) "Track" Events

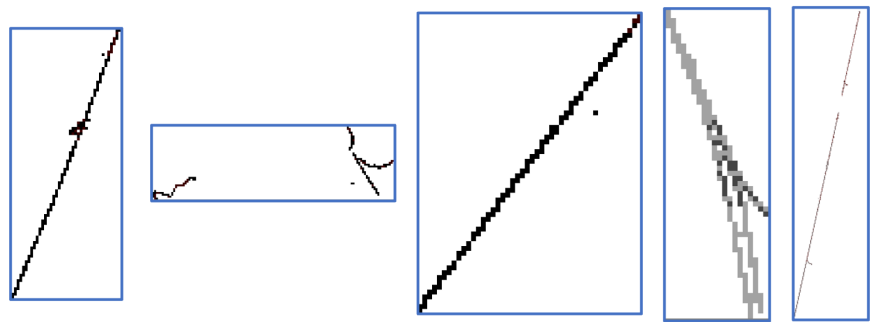

**Figure 6.** Examples of larger background events in the CdTe Eiger2 detector with a 6 keV threshold. See text for discussion. Displayed using the Albula software [2]. For the plots in (a) and (b) the number of counts in is printed in each pixel and grey pixels have just 1 or two counts, while colored pixels have many more.

events appear somewhat like a crater with a thin and steep wall. Examples are presented in figure 6a.

2. Mid-scale "Cluster" events where usually the aspect ratio of the distribution is not too large (say 1 to 5) and very often there is one pixel with a high (~100-1000) number of counts, while most pixels have 1-2 counts. Examples are presented in figure 6b.

3. "Track" events with a large aspect ratio (3 to >100) and, all pixels with only a few (always <10) usually 1-2 counts. Examples are presented in figure 6c.

Considering possible sources for these events, one speculates that the cluster events (2) may be due to alpha particles as may be from residual contaminants (e.g., radon) in the atmosphere. Alpha particles tend to be extremely energetic but stop easily so would lead to large energy deposition mostly in one pixel in the detector. The cluster may be created when alpha particle impacts one pixel and other particles scatter to nearby pixels. The track events (3) are consistent with what one expects for a CRM which deposits energy while travelling and also occasionally generates additional high-energy particles which create separate tracks. The crater events (1) are more difficult to understand. In particular, the high symmetry coupled with the tendency of a central pixel to have small counts seems hard to explain as being due to scattering events, so we speculate they might be due to a huge energy deposition in a pixel that then couples capacitively (i.e.: through the electronics) to the nearby pixels while saturating/freezing the pixel where the initial interaction occurred.

## 9. Time Slicing to Reduce the Effective Background Rate

When considering how to reduce the background for low-rate experiments, the practical question becomes how many of the background events are indistinguishable from the desired signal events. For example, if one knows the signal rate is low, then events of the type shown in figure 6 can safely be ignored, reducing the effective background rate. Extending this logic further, if the signal rate is low, and the frame rate is sufficiently large, then processing can generally be used to identify and then remove background events if one can somehow distinguish them from real signal events. There are different level of processing that can be considered, as will be addressed soon, however, it is clear that higher frame rates are generally desirable both to limit the loss in effective live time from the background and to help prevent misidentification of a multi-hit signal events as background. Thus, we refer to this general approach "time slicing" as we are slicing a longer exposure time (say 10s) into many frames (say 10,000 frames) to allow identification and removal of background events.

We start by considering a very simple case, assuming any frame at our 1 kHz frame rate with more than one illuminated pixel can be identified as a background event and ask what reduction in background rate might be possible if frames with those events are discarded. This rate is shown in figure 7 for the CdTe Eiger2 1M detector. The impact of such event selection is largest at lower thresholds, with, roughly, a factor of four improvement possible by throwing out all events with more than one pixel illuminated. As the threshold increases, the fraction of events with single pixels increases (see also figure 5) and the improvement becomes smaller. The fractional gains for such selection improve when 15 mm Pb shielding is added, as is consistent with an increasing fraction of multi-pixel CRM track events, as CRMs are generally only weakly reduced by this *relatively* thin Pb shielding. Thus, it is clear that if identifies and removes background events, significant reduction in the background rate may be possible.

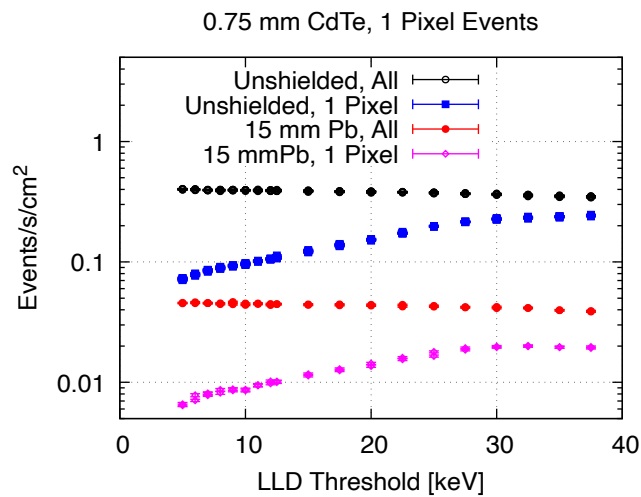

**Figure 7.** Rates for single pixel events as may be indistinguishable from signal in the 0.075 mm pixel CdTe Eiger2 1M detector. The data labeled "All" is the same as presented in figure 3 and is given here for reference.

Similar plots are shown in figure 8 for the Pilatus3 detectors, with the 0.172 mm pixel pitch. The improvement for the unshielded Si detector was about a factor of two, from ~0.08 to < 0.04 s/cm$^2$ as roughly half the events had more than one pixel illuminated. For the shielded silicon detector, where the shielding strongly suppressed the number of low pixel events (See figure 5c), the improvement is stronger, approximately a factor of 4 from ~0.018 to < 0.004 /s/cm$^2$.

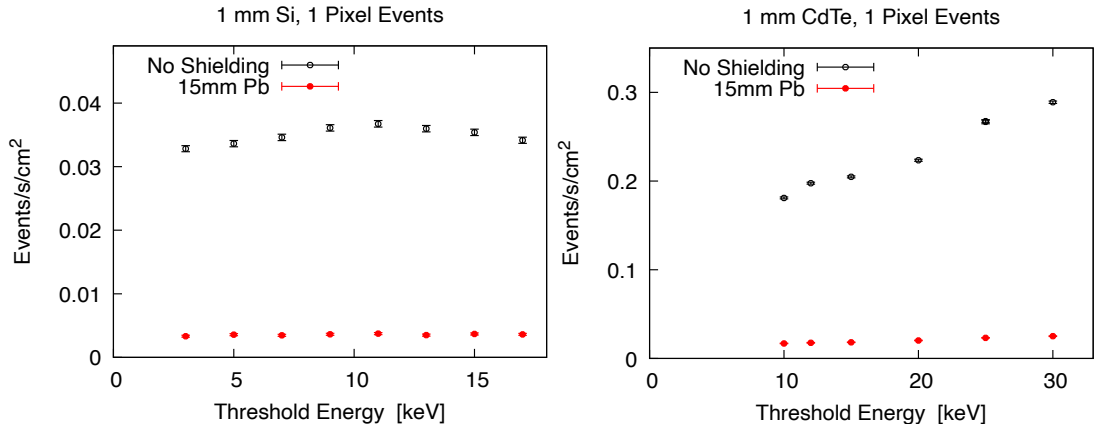

**Figure 8.** Rates of single pixel events in the Pilatus3 detectors with a 0.172 mm pixel pitch. See text for discussion.

We now consider somewhat more sophisticated processing, using the retriggering and/or an upper level on the Eiger2 detector. In particular, one can consider removing events that deposite more energy than the expected signal, e.g. are above some upper level, or removing events that lead to more than a single count in a pixel via the retriggering, or both. These results are shown in figure 9. In general, for such processing the upper level (ULD) is taken as 2.6 or 2.8 times the lower level (LLD) so that one is thinking in terms of detecting x-ray with an energy of double the lower discriminator threshold with the ULD set 30 or 40% above the nominal energy. e.g. for a 10 keV LLD, the ULD is 26 or 28 keV for an expected 20 keV photon energy. One notes that the retrigger, while behaving differently below 9 keV (as is not surprising given a CdTe detector is mostly aimed at higher energy photons) already affords signficant improvement in rate, especially for LLDs between 10 and 25 keV. However, the best results are obtained using the ULD, or, as is slightly better, using both the ULD and the retrigger. For example, considering as real background events only those that illuminate 1 pixel, and do not trigger the ULD, shows a background of <0.05/s/cm$^2$ is possible for an unshielded detector with a 15 keV LLD (so for ~30 keV photons) and <0.005/s/cm$^2$ if 15 mm of Pb shielding is used.

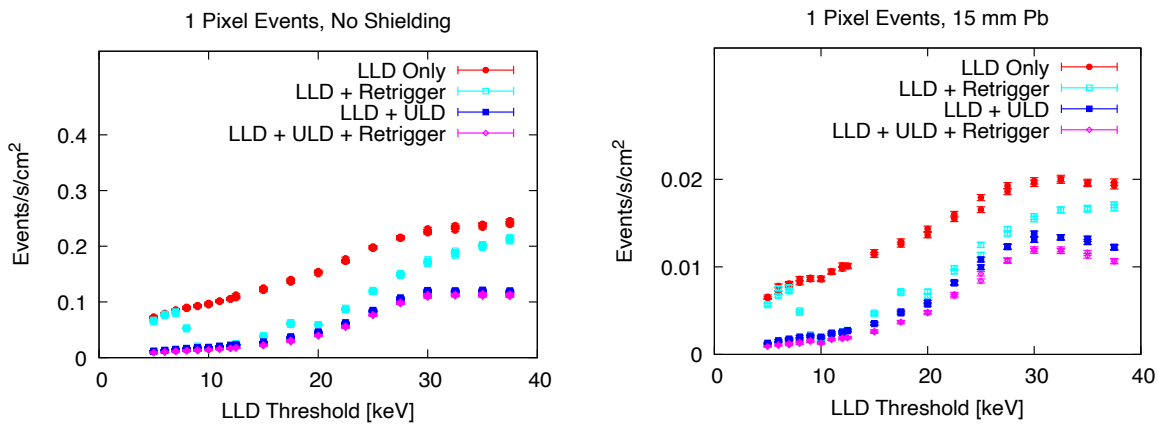

**Figure 9.** Background rates illuminating 1 pixel, processed in different ways for the Eiger2 CdTe 0.075 mm pixel pitch detector. The rates using both the ULD and retrigger is only slightly smaller than the result using the ULD alone so those points nearly overlap. See text for discussion.

## 10. Rate Correction for Time Slicing
The approach suggested in the previous section, where specific events are identified as background events, distinct from signal, and then ignored, has the potential to lead to systematic errors in data processing. At a minimum, it leads to an effective dead time in the region of the identified background event. Additionally, as signal rates increase, such processing may confound the estimate of the real signal rate since larger (muti-pixel) signal events may be removed as background. In particular, if we assume that every event that has more than one illuminated pixel in some area of the frame is a background event, then, as the signal rate increases there is an increasing chance that multiple illuminated pixels may be due only to signal events, which are then removed, effectively throwing away signal. This *both* reduces the effective live time of the detector *and* impacts the usual assumption that the mean measured rate is a good estimate of the mean signal rate: a correction is needed as the processing selects only single photon signal events. It is clear that these corrections are small if the frame rate is much larger than either the signal rate or the background rate, but as signal rate increases, these corrections can effectively limit the detector dynamic range. The purpose of this section is to estimate the magnitude of these effects assuming that the signal and background event rates are Poisson distributed and uncorrelated.

Several relations for Poisson statistics are given below. We assume that we focus on some area of the detector with an average of $\mu$ events in some time interval, T. Then the chance of measuring n events in that period is given by $P_T(n, \mu) = \mu^n e^{-\mu}/n!$ The first line below explicitly gives this distribution, indicates it is normalized, and that the average when summing over all events just will give $\mu$. The second line gives specific probabilities for $n = 0$ and $n = 1$ and the last line the probability for $n > 1$.

$$P_T(n,\mu) = \mu^n e^{-\mu}/n! \quad \sum_{n=0}^{\infty} P_T(n,\mu) = 1 \quad \sum_{n=0}^{\infty} n P_T(n,\mu) = \mu$$

$$P_T(n=0,\mu) = e^{-\mu} \quad P_T(n=1,\mu) = \mu e^{-\mu}$$

$$P_T(n>1,\mu) = 1 - e^{-\mu}(1+\mu) \approx \mu^2/2 \quad (\mu \ll 1)$$

The picture of the experiment is that we measure signal in an area of the detector for some time T, and remove all frames with identifiable noise counts, such as those with multiple illuminated pixels (e.g. with a track), or where the ULD fires, etc. This reduces the effective measurement time from T to $T_M$, by an amount that is exactly known during the experiment. As we are also throwing away all events that correspond to detecting multiple signal x-rays in the given area, we are in effect only measuring the rate of single photon, n=1, events. If we then measure N single events in a time $T_M$ the measured rate, M= N/ $T_M$ provides an estimate of $P_{T_M}(n=1,\mu) = \mu e^{-\mu}$ and not (directly) $\mu$ where $\mu$ is the combined average rate for single photon signal events and background events that are indistinguishable from single photon events. This can be inverted to give $\mu$ numerically, or using ProductLog or Lambert function. When $\mu$ and $M$ are small compared to unity, one has $\mu \approx M + M^2$. E.g. for $\mu$ =0.1 the correction will be about 10% of $\mu$.

We consider the magnitude of these corrections assuming, *momentarily*, a detection area of 1 cm$^2$ and a frame rate of 100 Hz or higher. The work above - which shows the background rate is always < 0.5/s/cm$^2$ - allows us to assume the average number of identifieable background events in a 10 ms frame will be $\mu_B < 0.005$. Thus the impact of the dead time from the indentifiable background events will be small and, more so, it is also known/measured. A more serious issue occurs as the signal rate increases. Assuming, for example, that we wish to keep the dead time < 10% then gives an upper bound on the allowable signal rate of $\mu \sim 0.1$, or a lower bound on the frame rate of about 10 times the signal rate. If the signal rate for an area of interest exceeds this rate, and we are discarding multi-pixel exposures, then the processing will begin to discard a quickly-increasing fraction of the frames. The increasing dead time will limit the efficiency of the measurement and, also the correction discussed above, needed to go from the average measured single photon rate to the real average signal rate, will become larger. While, in principle, these corrections (dead time and signal rate) are reasonably known, so one can probably tolerate values of $\mu \sim 0.3$, or perhaps more, such high rates will both limit the efficiency of the measurement as the dead time becomes larger, and also will make one increasingly sensitive to the reliability of the assumptions in the data processing (dead area, rigorous poisson statistics, etc). Thus, for safety, we generally aim to keep $\mu < 0.1$, or, the average signal rate in an area of interest < 10% of the frame rate. If we are then focussing on a 1 cm$^2$ area of a detector and collecting data with a 1 kHz frame rate, then one should insure the signal rate is comparable or less than about 100/s/cm$^2$. However, as we will discuss in the next section, the relevant area is rather smaller than 1 cm$^2$, so that higher signal rates, e.g. ~10$^4$/s/cm$^2$, can, in principle, reasonably be measured even while using time slicing.

We emphasize that the rates just mentioned, 100/s/cm$^2$ or 10$^4$/s/cm$^2$, are much larger than the maximum background rate measurred in the worst case in the present work (<0.5/s/cm$^2$) so that, depending on the experiment, it may in fact be easiest just to include multi-pixel events when estimating the average rate at high rates. How this might be done in detail will depend on the desired precision of the work and the desired dynamic range of the measurement. Also, as seen from the change in the statistical quality of the data in figure 12, one may want to also consider the influence of the relatively large and non-poissonian fluctuation in the number of illuminated pixels from CRM tracks. In other words, removing the CRM tracks both reduces the average background event rate and reduces the statistical fluctuations in the data – both should be considered before one relaxes the processing to allow multi-hit events.

## 11. Event Borders

The above discussion considers discarding a 1 cm$^2$ area when a background event is indicated. However, we only need to throw out that portion of the frame that has the background in it, and even the large background events of figure 6 illuminate much less than 1 cm$^2$ of area. See also the 3x3 cm$^2$ areas displayed in figure 1. Further, as mentioned above, a more important main limit on the detector dynamic range with the time slicing (assuming at least moderate, >~100 Hz, frame rates and that the real background rate is of the level discussed in section 6) is that multiple signal events may occur near eachother within a given frame and be mis-identified as a background event. Then, noting that signal events will illuminate only 1 pixel, it is clear the 1 cm$^2$ discussed above is much larger than is needed. However, in order to discuss this further, we need to specify how to distiguish background and signal events more precisely.

A straightforward extension of the discussion above is to assume that any illuminated pixel in a frame that has one count, no ULD firing, and is isolated from all other illuminated pixels is potential signal, and should be counted, while all other events are probably background and should be excluded. However, one then needs to set a border size to use to determine if a pixel is isolated. One would like to keep the border as small as possible to keep the excluded area small and the effective detector dynamic range large, but one would also like it large enough so that one does not misidentify part of a background event as signal. Pixels in background events can become isolated in at least two ways: flourescence and track fragmentation. If the LLD threshold is set below the flourescence energy, then, for CdTe, the fact that the attenuation length for flourescence x-rays can be ~120 μm (Section 3), means that a background event in one pixel may lead to detection of a flourescence event several pixels away. Fragmentation refers to the fact that as the LLD threshold is increased, it becomes increasingly possible that a CRM track can break up, fragment, into small separated segments (see figure 10) as the amount of charge distributed in each pixel is not uniform. This can lead to isolated pixels as the threshold is raised.

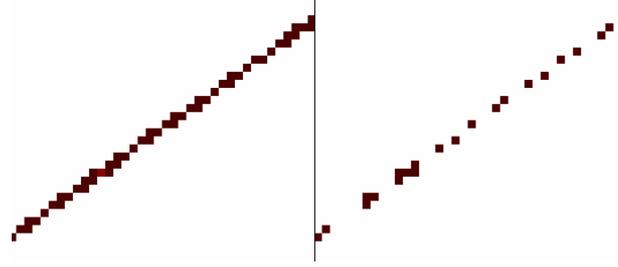

**Figure 10.** Event fragmentation in the Eiger2 CdTe detector. The left and right panels (divided by the vertical line) show the same CRM event with discriminator levels of 15 (left) and 29 keV (right) respectively. The fragmentation introduced by the higher discriminator threshold is clear.

The correct border size is potentially dependent on threshold, material, and pixel size – e.g. clearly from figure 10 the border should probably be set differently for different LLD settings. Therfore we look at the number of background events as a function of border for several different cases. Measured results are shown in figure 11, where we plot the number of isolated events for different border sizes as a fraction of the the events with a 1-pixel border. This should be monotonically decreasing, and the place where it levels off indicates where increasing the border becomes ineffective.

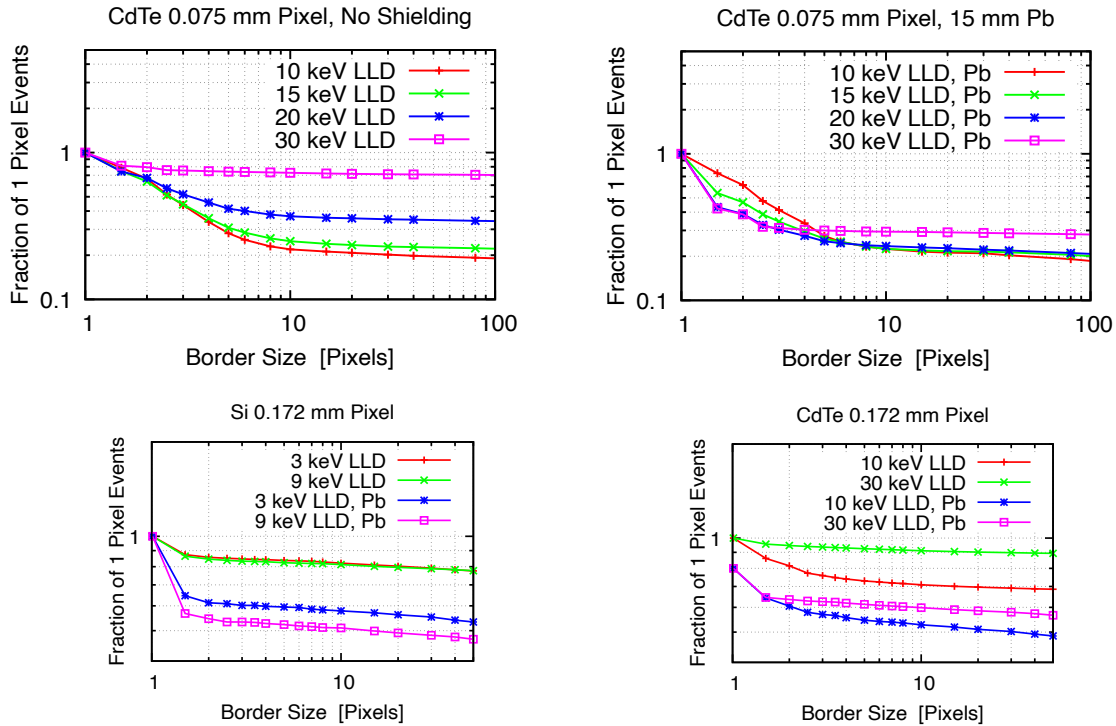

**Figure 11.** Impact of different border sizes on determining the number of isolated 1-pixel events. Note, for the 0.172 mm pixel CdTe sensor, the traces with Pb have been scaled by a factor of 0.8 for clarity in display. In all cases, "Pb" refers to the setup with 15mm Pb placed around the detector. See text for discussion

Considering first the 0.075 mm pixel size CdTe detector at lower thresholds and without shielding, it is clear that setting a border size of 8 pixels or 0.6 mm, so an effective event area of 1.13 mm$^2$ is enough to determine isolated pixels. Interestingly, the effective border size becomes smaller as the threshold increases, suggesting the dominant source of multi-pixel background events occurs via the CdTe fluorescence radiation and not track fragmentation. For safety, however, a border size of 8 pixels was used in the all the processing described above. The results for the other detectors are also shown. For the silicon detector, a border size of 2 pixels is mostly enough, while for the CdTe with larger pixel size, about 3 pixels are needed without shielding. However, for the larger pixel CdTe, in fact the impact of the border size is only about 30% total, while for the other detectors properly choosing the size of the border has a larger impact.

If we then return to the dicussion in section 10 about the effective limit on the dynamic range for increasing signal rates, we see that it is easy to to gain the indicated ~2 orders of magnitude. In particular, if we choose an 8 pixel border size for the CdTe Eiger2, corresponding to a 1.13 mm$^2$ signal event dead region, and set a conitnuous 1 kHz frame rate, then we will reach $\mu$ =0.1 at a signal rate of 100/s/1.13mm$^2$ or ~9000/s/cm$^2$. If achieving a larger singal rate is critical, or if higher energies and hence a higher LLD are used, then the border could be redcuced a factor of 2 or more, leading to $\mu$ =0.1 rates of >20,000/s/cm$^2$. Of course, this refers to the peak signal rate in any part of the detector, so requires some care, but, generally, for the low rate experiments that we are most concerned with, these are comfortable, even generous, limits.

## 12. Implementation and Example

The issues mentioned above are discussed rather generally, and it is useful to go through a definite example to show both a case where backgrounds are an issue, and how time-slicing improves the results. We use the CdTe Eiger2-1M detector delivered in 2022. It has a standard Detector Control Unit (DCU). The interface to the DCU was a Dell 7920 workstation (Xeon Silver 4215R 3.2GHz, 4 GHz burst, 8 Cores, 128GB memory) running Red-Hat Enterprise Linux (RHEL) with Python 3.9. DECTRIS' "stream2" interface was used to continuously collect dual threshold data at 1 kHz (0.001s frame duration). The data transfer from the DCU to the workstation was done using conventional copper-wire LAN cables & 1GB ethernet as this was adequate for our work. The frame processing was done via Python code which was imbedded into a "receiver" code, provided by DECTRIS, that was run on the 7920 workstation. The receiver code operates with several (typically 8) parallel processes via the Python multiprocessing package to keep up with the data rate. Our frame processing code (performing steps 0-4 below) was imbedded in the receiver code so could take advantage of the parallelization. The top level user interface was through the SPEC control program (Certified Scientific Software [15]) running on an iMac. Socket I/O was used in SPEC to communicate with a Python server running on the 7920 workstation.

The test experiment was a very low-rate measurement using high (sub-meV-) resolution inelastic x-ray scattering (see [16] for a general introduction to the field). In these experiments one collects data by doing repeated scans over some range of parameters: at each point in the scan conditions are held stable while one collects data (counts) for a fixed time period, 10s in the present case, then conditions are changed, and one counts again. A typical scan contains 500 to 1000 points and requires 2-3 hours. Several scans are usually made and summed, possibly with binning, to improve the statistical quality of the data. In the present case, the setup had an extremely low signal rate and we integrated for 30 hours. At each 10s point in each scan, we collected 2x10,000 x 0.001 s frames of time-sliced data (10,000 images for each of the LLD and ULD) from the detector. All empty frames (typically > 95% of the frames were empty) were noted and discarded, and the remaining frames were processed over a region of interest of 178x737 pixels. For each non-empty frame pair, the positions and values for all pixels with counts for the LLD, the ULD were saved in a text file. Also saved to that file were a "selected" set of pixels corresponding to the processing to be described below. Then, at the end of the 10s counting time, all the individual files for a given scan point were combined into one file for that scan point. The data was sparse enough so that result from the 20k measured frames of 1M pixels could be placed into a file of <1kbyte in size. One notes that the data was saved in such a way that each frame within that 10s was distinct, so that the complete data for the 10s count time was saved without losses, except for the time ordering of the frames. For each scan we also usually saved one DECTRIS standard format "cbor" file with the detector information – thresholds, masks, etc. That file was large (~17 MB) and sometimes deleted when considered redundant.

The selected pixels were determined from the pixels with counts by removing all events that were distinguishable as background. In detail, our data processing consisted of the following steps for each 0.001 s frame pair with counts:
0. Select (make a list, positions and values, of) all illuminated pixels from the LLD frame.
1. Remove all pixels that are within a distance of 8 pixels of another pixel.
2. Remove all pixels with >1 count in the pixel (e.g., using the retriggering).
3. Remove all pixels where the ULD image has counts, leading to a final list of "selected" pixels.
4. Save the illuminated LLD, ULD and selected pixels to a text file.

This then usually led to a short list of selected pixels, typically not more than 1 or 2, for each frame pair.

We say a few more words about the data processing and collection, as may be useful for those interested in doing similar work. The initial trigger of the acquisition at each scan point required about 0.3 s, as was dominated by the detector control software response time (this might be reduced in the future using hardware triggering). The slowest single step in the frame processing was the selection of the illuminated pixels, step 0, as implemented by the numpy.where() command, as examining a large number of pixels takes some time. Tests suggested dedicated compiled code might be faster by about a factor of 2. However, over the 178x737 pixel region of interest in the experiment, the numpy.where() command was sufficiently fast. The processing to determine the distance between the pixels in a frame (step 1) was slow for frames with many illuminated pixels, as, in our simple implementation, it scaled as the square of the number of illuminated pixels. However, as the number of illuminated pixels was generally not large this did not limit our work, and so was not heavily optimized. We used socket polling via the Python zmq package to share information between the control sever and the independent asynchronous multi-process receiver code. An initial version of the sever returned a lot of information to the controlling SPEC software, including all selected pixels. However, this was superseded by a version that returned more limited information (filenames, integrals, number of illuminated pixels) and relied on saving the image information on the 7920 server, with, as noted above, one ~ kbyte text file holding the data from all the images for each 10s point in the scan. The total overhead in the scan for triggering the data collection, collecting dual threshold data at 1 kHz, processing and saving the data, returning values to SPEC, and for SPEC to read many other counters, was typically 1 to 1.2 s. While not optimal, this 10 to 12% deadtime on our 10s counting time was acceptable for the present work.

The measurement was done on a sample of liquid water in ambient conditions at BL43LXU [17] of SPring-8. The response of this sample is expected to be a peak at zero energy transfer with weak shoulders, as can be seen, e.g., in [18]. In the present measurement we used a 25.7 keV x-ray energy corresponding to the Si(13 13 13) back reflection. We collected data using 13.4 x 55.4 mm$^2$ region of the Eiger2 detector with thresholds set at 12.8 and 33 keV. The detector was partly shielded with lead, 5 to 10 mm, and a tungsten plate, all placed as to reasonably fit into our space-constrained temporary test setup. The scattering was collected over 30 hours of scanning, and then processed to relate the position in the detector and other scan parameters to a reduced detector coordinate proportional to energy transfer. The results with and without time slicing are shown in figure 12 and the reduction of the uniform background is clear. Further, the statistical quality of the data when the time slicing is used is much better. That is because the number of pixels illuminated by cosmic ray muons is highly variable, depending on the exact trajectory of the CRM through the detector. Thus, even in this temporary test setup, the value of the time slicing is clear in both the strong reduction of the uniform (energy-independent) background and the improvement in the statistical quality of the data.

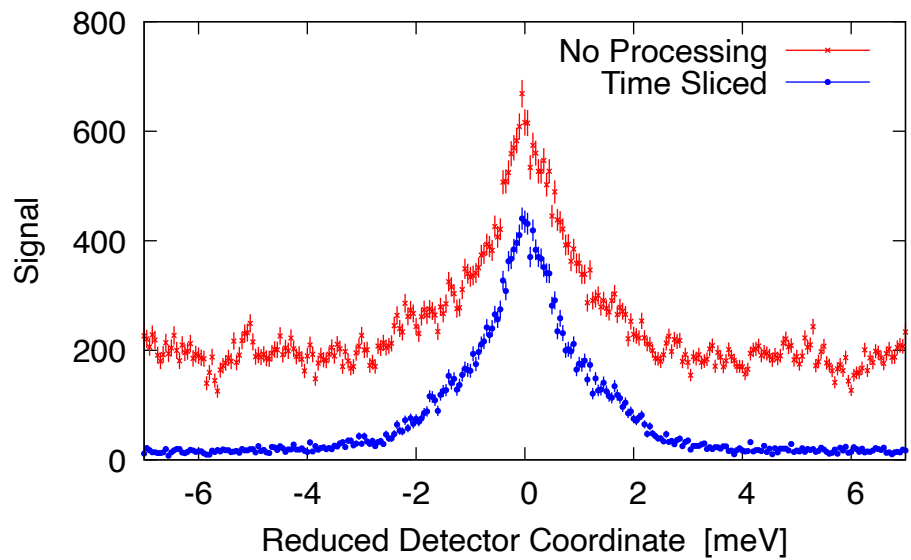

**Figure 12.** Test measurement of a water sample for 30 hours using the Eiger2 CdTe detector. The signal is the peak about x=0. The reduction of the background and improvement in data quality when time slicing is used is very clear. Error bars correspond to the square-root of the number of counts in a bin.

## 13. Summary and Discussion

We have quantified the background rates in several pixel array detectors, demonstrating event rates as indicated in table III. In general, adding 15 mm of Pb shielding is effective to reduce the rates by a factor of 4 to 8, consistent with the largest background source for an unshielded detector being gamma rays from decay of trace amounts of unstable isotopes present in the environment. The background rates observed in a silicon sensor are smaller than (1/5 to ½ of) those observed with a CdTe sensor, consistent with the higher stopping power of the CdTe for the gamma ray background.

|  | Measured Events/s/cm$^2$ | | |
| --- | --- | --- | --- |
|  | No Processing | Processed with Time Slicing | |
|  |  | Single Pixel Events | Single Pixel, ULD (& Retrigger) |
| 0.75 mm CdTe No Shielding | ~0.4 (LLD < 40keV) <br> ~0.2 (LLD 75 keV) | 0.12 (15 keV LLD) <br> 0.22 (40 keV LLD) | < 0.025 (LLD<15 keV) <br> ~0.1 (30<LLD<40 keV) |
| 0.75 mm CdTe 15 mm Pb | < 0.05 | 0.012 (15 keV LLD) <br> 0.02 (40 keV LLD) | < 0.003 (LLD<15 keV) <br> ~0.013 (30<LLD<40 keV) |
| 1 mm Si No shielding | <0.08 | < 0.035 | - |
| 1 mm Si 15 mm Pb | <0.02 | < 0.004 | - |

Table III. **Summary of Measured Background Rates.** The rates for the 1 mm thick CdTe Pilatus3 detector are similar but slightly higher than for the 0.75 mm thick Eiger2 detector (see sections 6 & 9)

Time slicing a longer measurement can be used to help identify and remove background events, with additional reductions of background rates of factors of 4 to 10 realizable beyond those above, depending on detailed conditions. This approach reduces the effective dynamic range of the detector, but should still, conservatively, allow local signal rates ~10,000/s/cm$^2$ for kHz slicing (kHz frame rates) if a correction is applied. While this limit should not be ignored, it also is probably not serious as the conditions where the background levels discussed here (<0.5/s/cm$^2$) will have the largest impact will generally be lower-rate measurements. We note that collecting an upper level (ULD) image and then discarding high-energy events was effective. Finally, using both shielding and time slicing with processing, we were able to reduce background rates to ~3-4 /1000s/cm$^2$, for thresholds <15 keV (photon energies <30 keV) which allows one to consider rather low-rate experiments. The value of this processing was demonstrated in a test experiment.

We briefly discuss some options that might be considered for further reducing backgrounds. Of course, one can increase the thickness of the Pb shielding. However, our measurement showed a factor of four reduction in background for the first 5 mm of Pb, and only an additional factor of two for the next 10 mm with a CdTe sensor. This suggests significant further reduction will require much more shielding, which, with a Pb density of 11.3 g/cc, can become very heavy. A third discriminator threshold setting might also allow further background reduction, as is suggested by the decrease in the effectiveness of the processing in the CdTe as the LLD increases (figures 7-9): setting a putative third threshold to a minimum value could facilitate identification of track events, especially when the main target is detecting higher energy x-rays (see figure 10). In general, if backgrounds are a major concern, it is also clear one should choose a sensor material and thickness that is optimized for the signal energy of interest, and *not* higher energy: one would like poor efficiency for the generally high-energy gamma background. This might make it interesting to consider a GaAs sensor when the goal is detecting x-ray energies in the 20-50 keV range as GaAs will generally have lower chance of detecting the mostly high (>100keV) energy gammas in the background than CdTe. One also speculates that the lower fluorescence energies, <12 keV, in GaAs, as opposed to 23-31 keV for CdTe, might improve the response of the detector. Finally, we note that one can consider more sophisticated processing schemes (e.g., track identification) but those probably are most interesting in special cases.

## Acknowledgements
We are grateful to the staff of DECTRIS for their kind patience answering questions about these detectors, helping with some of the software control, for their comments on the paper, and for loan of a detector that was used for some initial tests. In particular, we thank Takeyoshi Taguchi (DECTRIS Japan) Sascha Grimm, Tilman Donath, and Clemens Schulze-Briese (DECTRIS). AB is grateful to Bixente Rey for help with the SPEC server code, and thanks T. Oguchi for the Pb shielding plates. The work was performed in the Materials Dynamics Laboratory of the RIKEN SPring-8 Center.

## Author Contributions
AB set the scope of the project, performed the off-line measurements, processed the data and wrote the paper. DI helped with specifications for equipment procurement, contributed to some SR measurements, participated in the test measurement (section 12) and offered comments on the paper.